\documentclass[letterpaper, 10pt, conference]{ieeeconf}      % Use this line for a4
\IEEEoverridecommandlockouts                              % This command is only
\usepackage{graphicx}
\usepackage{amsmath}
\usepackage{amssymb}
\usepackage{float}

\newtheorem{definition}{Definition}
\newtheorem{remark}{Remark}
\newtheorem{problem}{Problem Statement}
\newtheorem{motivation}{Motivation}
\newtheorem{assumption}{Assumption}

\title{\LARGE \bf
Formation of Multiple Groups of Mobile Robots Using Sliding Mode Control
}

\author{Soumic Sarkar$^{1}$ and Indra Narayan Kar$^{2}$% <-this % stops a space
\thanks{$^{1}$S. Sarkar is with Department of Electrical Engineering, Indian Institute of Technology. Delhi, New Delhi-110 016, INDIA
        {\tt\small soumic4it at gmail.com}}%
\thanks{$^{2}$I.N. Kar is with the Faculty of Electrical Engineering, Indian Institute of Technolog. Delhi, New Delhi-110 016, INDIA
        {\tt\small ink at iit.ac.in}}%
}

\begin{document}

\maketitle
\thispagestyle{empty}
\pagestyle{empty}

%%%%%%%%%%%%%%%%%%%%%%%%%%%%%%%%%%%%%%%%%%%%%%%%%%%%%%%%%%%%%%%%%%%%%%%%%%%%%%%%
\begin{abstract}
Formation control of multiple groups of agents finds application in large area navigation by generating different geometric patterns and shapes, and also in carrying large objects. In this paper, Centroid Based Transformation (CBT) \cite{c39}, has been applied to decompose the combined dynamics of wheeled mobile robots (WMRs) into three subsystems: intra and inter group shape dynamics, and the dynamics of the centroid. Separate controllers have been designed for each subsystem. The gains of the controllers are such chosen that the overall system becomes singularly perturbed system. Then sliding mode controllers are designed on the singularly perturbed system to drive the subsystems on sliding surfaces in finite time. Negative gradient of a potential based function has been added to the sliding surface to ensure collision avoidance among the robots in finite time. The efficacy of the proposed controller is established through simulation results. 
\end{abstract}

%%%%%%%%%%%%%%%%%%%%%%%%%%%%%%%%%%%%%%%%%%%%%%%%%%%%%%%%%%%%%%%%%%%%%%%%%%%%%%%%
\section{INTRODUCTION}

The study on the collective behaviour of birds, animals, fishes, etc. has not only drawn the attention of biologists, but also of computer scientists and roboticists. Thus several methods of cooperative control \cite{c10} of multi-agent system have evolved, where a single robot is not sufficient to accomplish the given task, like navigation and foraging of unknown territory. These methods can broadly be categorized as the behaviour based approach (\cite{c1}-\cite{c3}), leader follower based approach \cite{c4}-\cite{c5}, virtual structure based approach \cite{c6}-\cite{c9}, artificial potential based navigation \cite{c13}-\cite{c15}, graph theoretic method \cite{c11}-\cite{c12}, formation shape control \cite{c16}-\cite{c21}. Among other works carried out on single group of robots, cluster space control \cite{c34}, distance based formation \cite{c35}, formation control of nonholonomic robots \cite{c4}, kinematic control \cite{c27}, and mobile robots subject to wheel slip \cite{c36}, segregation of heterogeneous robots \cite{c37}, are to name a few.\\
The problem associated with the formation control of multi-agent system is that it becomes difficult to accurately position the robot within the group, as the number of robots increases. To address this issue shape control and region based shape control have been proposed, such that the robots form a desired shape during movement. The desired shape can be union or intersection of different geometric shapes. Region based shape control have been extended to multiple groups of robots \cite{c22}-\cite{c24}. However, the robots can stay anywhere inside the specified region without colliding with each other. This means that the position of a robot inside a group can be specified and can further be controlled. Therefore
the position of a group of robots inside a larger group of robots can also be specified and controlled. Moreover, when it comes to the control of multiple groups of robots, there should be at least one robot to convey the information of that group to another group.\\
In an attempt to solve the positioning accuracy, we propose a hierarchical topology, here in this paper, which is based on the centroid based transformations \cite{c16}-\cite{c19} for single group of robots. In this architecture, the large group of robots have been partitioned into relatively small basic units containing three or four robots. Then the centroid of each unit have been connected to form larger module containing more robots. Extending the process will give a hierarchical architecture which is a composition of relatively smaller modules. As the construction of this topology involves connecting the centroid, it has been named Centroid Based Topology (CBT). The CBTs basically capture the constraint relationship among the robots. Using CBT it is possible to separates shape variables from the centroid and this technique separates the formation shape controller and tracking controller design. As the centroid moves, the entire structure moves maintaining the shape specified by the shape variables. In this paper, we study the formation of multiple groups of robots in a modular architecture. Using the concept of CBT, we define \textbf{intra group shape variables}, \textbf{inter group shape variables} along with centroid. Based on this modular structure, sliding mode based controllers have been designed for each module. The gains of the controllers have been chosen such that the subsystems reach the sliding surface at different time featuring the stretched time scale properties of singularly perturbed system. Singular perturbation based sliding mode controller design gives us the freedom to run the superfast [intra group formation] to slowest dynamics [tracking of centroid] simultaneously without waiting for the convergence of others. Furthermore, potential function based sliding surfaces have been selected to design controllers to avoid inter robot collision in finite time.

\section{Main Result}
Suppose that there are $m$ groups of $n$ robots in a plane. Define a set $\{n_i:\sum_{i=1}^mn_i=n\}$ with $i=1,...,m$ to denote the number of robots in $i^{th}$ group. Let $p_{ij}=[x_{ij},y_{ij}]^T\in\mathbb{R}^2$, $i=1,...,m$ and $j=1,...,n_i$ denote the position of $j^{th}$ robot in $i^{th}$ group. Suppose that each robot is governed by the following dynamics
\begin{equation}
\label{801}
\begin{split}
\ddot{p}_{ij}&=A_{ij}(\theta_{ij},\dot{\theta}_{ij})\dot{p}_{ij}+B_{ij}(\theta_{ij})u_{ij}+C_{ij}(\dot{\theta}_{ij}) \\
J_{ij}\ddot{\theta}_{ij}&=\frac{R_{ij}}{r_{ij}}(\tau_{rij}-\tau_{lij})
\end{split}
\end{equation}
where
$$
A_{ij}(\theta_{ij},\dot{\theta}_{ij})=\begin{bmatrix}
-\sin\theta_{ij} \cos\theta_{ij} \dot{\theta}_{ij} & -\sin^2\theta_{ij} \dot{\theta}_{ij} \\
\cos^2\theta_{ij} \dot{\theta}_{ij} & \sin\theta_{ij} \cos\theta_{ij} \dot{\theta}_{ij}
\end{bmatrix}
$$
\[
B_{ij}(\theta_{ij})=\begin{bmatrix}
\frac{\cos\theta_{ij}}{m_{ij}r_{ij}}-\frac{d_{ij}R_{ij}\sin\theta_{ij}}{J_{ij}r_{ij}} & \frac{\cos\theta_{ij}}{m_{ij}r_{ij}}+\frac{d_{ij}R_{ij}\sin\theta_{ij}}{J_{ij}r_{ij}} \\
\frac{\sin\theta_{ij}}{m_{ij}r_{ij}}+\frac{d_{ij}R_{ij}\cos\theta_{ij}}{J_{ij}r_{ij}} & \frac{\sin\theta_{ij}}{m_{ij}r_{ij}}-\frac{d_{ij}R_{ij}\cos\theta_{ij}}{J_{ij}r_{ij}}
\end{bmatrix}
\]
$$
C_{ij}(\dot{\theta}_{ij})=\begin{bmatrix}
-d_{ij}\dot{\theta}_{ij}^2\cos\theta_{ij} \\
-d_{ij}\dot{\theta}_{ij}^2\sin\theta_{ij}
\end{bmatrix}
$$
where, $m_{ij}$ is the mass, $J_{ij}=I_{ij}-m_{ij}d_{ij}^2$, $I_{ij}$ is moment of inertia, $R_{ij}$ is the distance between left and right wheels, $r_{ij}$ is the radius of each wheel, $d_{ij}$ is the distance from wheel axis to the center of mass, and $\theta_{ij}$ is the orientation and $u_{ij}=[\tau_{rij},\tau_{lij}]^T\in\mathbb{R}^2$ is the control torque input of $j^{th}$ robot in $i^{th}$ group respectively. We use the notations $A_{ij}$, $B_{ij}$ and $C_{ij}$ to denote $A_{ij}=A_{ij}(\theta_{ij},\dot{\theta}_{ij})$, $B_{ij}=B_{ij}(\theta_{ij})$ and $C_{ij}=C_{ij}(\dot{\theta}_{ij})$ for $i=1,...,m$ and $j=1,...,n_i$. We write the combined dynamics of $n$ robots in augmented form as
\begin{equation}
\label{eq23}
\ddot{X}=\mathbf{A}\dot{X}+\mathbf{B}U+\mathbf{C}
\end{equation}
where, $X=[p_{11}^T,...,p_{nn_m}^T]^T$ $\mathbf{A}=diag\{A_{11},...,A_{nn_m}\}$, $\mathbf{B}=diag\{B_{11},...,B_{nn_m}\}$, $\mathbf{C}=diag\{C_{11},...,C_{nn_m}\}$ and $U=[u_{11}^T,...,u_{nn_m}^T]^T$.\\
Before presenting our results, we provide a few definitions. We first define shape vectors for single group of robots and then using that we define shape vectors for multiple groups of robots.
\begin{definition}
Let $q_i=[x_i,y_i]^T\in\mathbb{R}^2, i=1,...,n$ denote the positions of a system of $n$ particles with respect to fixed inertial coordinate frame of reference. Let there be another coordinate system $\xi=[\xi_i,\xi_c]^T\in\mathbb{R}^{2n}, i=1,...,n-1$, where, $\xi_i=\sum_{i,j=1}^{n}p_{ij}z_i\in \mathbb{R}^{2n}$ are \textit{shape vectors}, where $p_{ij}\in\mathbb{R}$, and are not all $0$s, and $\xi_c=\frac{1}{n}\sum_{i=1}^n q_i\in\mathbb{R}^2$ denotes the \textit{centroid} of all positions. Then we define a real linear mapping $\mathbb{R}^{2n}\times\mathbb{R}^{2n}:\mathbb{R}^{2n}\rightarrow\mathbb{R}^{2n}$ as
\begin{equation}
\label{rtrans}
\Phi:q\rightarrow\xi
\end{equation}
Specific applications of such mapping $\Phi$, $\mathbb{R}^{2n}\times\mathbb{R}^{2n}:\mathbb{R}^{2n}\rightarrow\mathbb{R}^{2n}$ can be found in \cite{c17} and \cite{c19}. For brevity we call the mapping $\Phi$ as \textit{Centroid Based Transformation (CBT)} for single group of robots. 
\end{definition}
One example of CBT is \textit{Jacobi transformation} to get \textit{Jacobi vectors} for \textit{Jacobi shape space}\cite{c18}. We will consider \textit{Jacobi vectors} as an example, to deduce our results in this article, though the results will comply similarly with other CBTs \cite{c19}. We give more stress to defining a new coordinate system to analyse the behaviour of the particles with respect to that reference frame, rather than investigating an interaction topology (communication among the agents), as in Graph theory \cite{c10}, with respect to specific coordinate system (Cartesian coordinate or Inertial frame).
\begin{definition}
Let there be another coordinate system $Z=[Z_s^T,Z_r^T,z_c^T]^T\in\mathbb{R}^{2n}$, where $Z_s\in\mathbb{R}^{2(n-m)}$ are \textbf{Intra Group Shape Vectors}, $Z_r\in\mathbb{R}^{2(m-1)}$ are \textbf{Inter Group Shape Vectors} and $z_c=\frac{1}{n}\sum_{1=1,j=1}^{m,n_i}p_{ij}\in\mathbb{R}^{2n}$ is the overall \textbf{Centroid} of all the robots. Let $Z_s=[Z_1^T,...,Z_m^T]^T$ with $Z_i\in\mathbb{R}^{2(n_i-1)}$, $i=1,...,m$ denote the intra group shape vectors of $i^{th}$ group of robots, then $Z_i=[z_{i1}^T,...,z_{i(n_i-1)}^T]^T$ for $i=1,...,m$ with $z_{ik}=\sum_{i=1,j=1}^{m,n_i}a_{ij}p_{ij}\in\mathbb{R}^2$ denote $k^{th}$ shape vector in $i^{th}$ group, where $a_{ij}\in\mathbb{R}$ for $k=1,...,n_i-1$, $i=1,...,m$. Let the centroids of $m$ groups be denoted by $\mu_1,...,\mu_m\in\mathbb{R}^2$ with $\mu_i=\frac{1}{n_i}\sum_{j=1}^{n_i}p_{ij}$, $i=1,...,m$. Define $Z_r=[z_{r1}^T,...,z_{r(m-1)}^T]^T$ with $z_{rk}=\sum_{i=1}^{m}b_{i}\mu_i$, $k=1,...,m-1$ where $b_i\in\mathbb{R}$. We then define a linear mapping $\mathbb{R}^{2n\times2n}:\mathbb{R}^{2n}\rightarrow\mathbb{R}^{2n}$ as
\begin{equation}
\label{multitrans}
\Phi_M:X\rightarrow Z
\end{equation}
We call $\Phi_M$, CBT for multiple groups of robots. The matrix $\Phi_M$ can also be written as 
\begin{equation}
\label{brkmtrans}
\Phi_M=\left[\Phi_1^T,\hdots,\Phi_m^T,\Phi_{r}^T,\Phi_{c}^T\right]^T
\end{equation}
where $\Phi_i\in\mathbb{R}^{2(n_i-1\times n)}$, $i=1,..m$, $\Phi_{r}\in\mathbb{R}^{2((m-1)\times n)}$, $\Phi_{c}\in\mathbb{R}^{2(1\times n)}$ correspond to the coefficient matrix associated with the intra and inter group shape vectors $Z_1,...,Z_m$, $Z_r$ and the centroid $z_c$ respectively.
\end{definition}
\begin{remark}
Note that CBT for multiple groups of robots is deduced by hierarchical application of CBT for single group of robots which is detailed in \cite{c39}-\cite{c41}.
\end{remark}
Using the transformation $\Phi_M$, \eqref{eq23} can be written as
\begin{equation}
\label{eq25}
\ddot{Z}=\mathbf{P}\dot{Z}+F+\mathbf{R}
\end{equation}
where $\mathbf{P}=\Phi_M \mathbf{A} \Phi_M^{-1}; \ F=\Phi_M \mathbf{B} U; \ \mathbf{R}=\Phi_M \mathbf{C}$. 
\begin{assumption} 
It is assumed that the robots are capable of communicating with each other and there should be at least one robot in the entire group with high communication and computation overhead [to calculate the centroid from the positional information communicated by all the robots].
\end{assumption}
\begin{problem}
Let the desired vectors in the transformed domain as $Z_d=[Z_{sd}^T,Z_{rd}^T,z_{cd}^T]^T\in\mathbb{R}^{2n}$, where $Z_{sd}=[Z_{1d}^T,...,Z_{md}^T]^T\in\mathbb{R}^{2(n-m)}$ is desired intra group shape vector with $Z_{id}\in\mathbb{R}^{2(n_i-1)}$, $i=1,...,m$, being the desired shape vector of $i^{th}$ group of robots. The error $Z_e=Z-Z_d$ is defined similarly as $Z_e=[Z_{se}^T,Z_{re}^T,z_{ce}^T]^T\in\mathbb{R}^{2n}$, where $Z_{se}=[Z_{1e}^T,...,Z_{me}^T]^T\in\mathbb{R}^{2(n-m)}$ is the intra group shape error, with $Z_{ie}=Z_i-Z_{id}$ is the shape error of $i^{th}$ group, $Z_{re}=Z_r-Z_{rd}$ is the inter group shape error, and $z_{ce}=z_c-z_{cd}$ is the tracking error. Define a set of time scales $t_s,t_r,t_c\in\mathbb{R}$ with $t_s=\frac{t}{\epsilon_1\epsilon_2}$, $t_r=\frac{t}{\epsilon_1}$. The scalars $\epsilon_i$, $i=s,r$ are such chosen that $t_s<t_r<t_c<t$, $t$ being the total time of operation. Then the problem statement can be described as to design $F$ in \eqref{eq25} such that $\lim_{t\rightarrow t_i}Z_{ie}=0$, $i=1,...,m$, $\lim_{t\rightarrow t_r}Z_{re}=0$ and $\lim_{t\rightarrow t_c}z_{ce}=0$.
\end{problem}
\begin{motivation}
The motivation of this work can be best clarified with the example given in the following diagram
\begin{figure}[h]
\centering
\includegraphics[width=8cm, height=4cm]{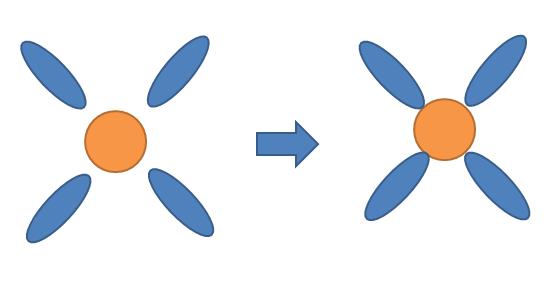}
\caption{Petal formation}
\label{fig:petal}
\end{figure}
Suppose that we want a flower-like formation of Fig. \ref{fig:petal}. We want the core and the petals to come to formation first [intra group formation] simultaneously with slower step toward the petals joining the core [inter group formation] and with a more slower step toward the tracking of the centroid of formation to a given trajectory. This could be treated as an example of three time scale convergence approach. Singular perturbation based controller design gives us the freedom to run the superfast [intra group formation] to slowest dynamics [tracking of centroid] simultaneously without waiting for the convergence of others.
\end{motivation}

\subsection{Sliding mode controller design in three time scale and stability analysis}

As given in our previous work \cite{c39}, the matrix $\mathbf{P}$ and $\mathbf{R}$ of equation \eqref{eq25} in subsection A, can be written in the following form
$$
\mathbf{P}=\left[
\begin{array}{c}
\mathbf{P}_s \\ \hline
\mathbf{P}_r \\ \hline
\mathbf{P}_c
\end{array}\right] \ ; \
\mathbf{R}=\left[
\begin{array}{c}
\mathbf{R}_s \\ \hline
\mathbf{R}_r \\ \hline
\mathbf{R}_c
\end{array}\right]
$$
Therefore, the collective dynamics of \eqref{eq23} can be separately written in the form of intra group shape dynamics $(Z_s)$, as follows
\begin{equation}
\label{eq101}
\ddot{Z}_s=\mathbf{P}_s \dot{Z}+F_s+\mathbf{R}_s
\end{equation}
where, $
Z_s=\mathbf{\Phi_m} X; \ F_s=\mathbf{\Phi_m} BU
$. 
The inter group shape dynamics $(Z_r)$ is written as,
\begin{equation}
\label{eq102}
\ddot{Z}_r=\mathbf{P}_r \dot{Z}+F_r+\mathbf{R}_r
\end{equation}
where, $
Z_r=\Phi_r X; \ F_r=\Phi_r BU
$. 
The dynamics of the centroid $(z_c)$ is expressed as,
\begin{equation}
\label{eq103}
\ddot{z}_c=\mathbf{P}_c \dot{Z}+f_c+\mathbf{R}_c
\end{equation}
where, $z_c=\Phi_c X; \ f_c=\Phi_c BU$. \\
\textit{1) Control law for centroid:} The controller that manages the centroid to track the given trajectory, is designed to be the last to converge to the desired value. The switching surface for the centroid dynamics is defined by
\begin{equation}
\label{surfc}
s_c(t)=cz_{ce}+\dot{z}_{ce}
\end{equation}
The equivalent control law (setting $\dot{s}_c=0$)  is given by
\begin{equation}
\label{cntrlc}
u_{c_{eq}}=-c\dot{z}_{ce}-\textbf{P}_c\dot{Z}-\mathbf{R}_c+\ddot{z}_{cd}
\end{equation}
The control law \eqref{cntrlc} only takes the system trajectory towards the origin along the sliding surface \eqref{surfc}. But trajectories which do not initiate on the sliding surface is required to reach the surface so that they can slide along the surface towards the origin. The following control law satisfies the reachability condition.
\begin{equation}
\label{reachc}
\bigtriangleup u_c=-\delta_csgn(s_c)
\end{equation}
where, $\delta_s$ is a positive scalar.
%$$
%sig(s)=\frac{1-e^{-\mid s\mid}}{1+e^{-\mid s\mid}}\geq 0 \ \ 
%$$
$$
sgn(s)=\begin{cases}
1 \ for \ s>0 \\
-1 \ for \ s<0
\end{cases}
$$
Then the sliding mode control law for centroid dynamics can be written as
\begin{equation}
\label{totcntrlc}
f_c=u_{c_{eq}}+\bigtriangleup u_c
\end{equation}
\textit{Theorem 1} The control law \eqref{totcntrlc} will asymptotically stabilize the subsystem \eqref{eq103} in finite time.\\
\textit{Proof:} Define a Lyapunov function for the subsystem \eqref{eq103} as
\begin{equation}
\label{centliap}
V(s_c)=\frac{1}{2} s_c^Ts_c
\end{equation}
The time derivative of \eqref{centliap} gives
\begin{equation}
\dot{V}(s_c) = -\delta_cs_csgn(s_c) =-\delta_c\mid s_c \mid\leq-\delta_cV(s_c)^\frac{1}{2}
\end{equation}
As $\dot{V}(s_c)+\delta_cV(s_c)^\frac{1}{2}\leq 0$ it follows from Theorem 1 that the sliding surface $s_c$ is finite time stable and there exist a finite time $t_c\leq \frac{2}{\delta_c}V(s_c(0))^{\frac{1}{2}}$ such that $z_{ce}\rightarrow 0$ for all $t\geq t_c$.\\
\textit{2) Control law for inter group shape dynamics:} To serve the purpose of different time scale convergence, the control law of \eqref{eq102} is chosen that the error dynamics is written in the form of singularly perturbed system as
\begin{equation}
\label{fast1}
\epsilon_1\ddot{Z}_{re}=-\delta_r sgn(s_r)
\end{equation}
where $s_r$ is the sliding surface for the dynamics of \eqref{eq102} and is chosen to be
\begin{equation}
\label{surfr}
s_r(t)=rZ_{re}+\dot{Z}_{re}
\end{equation}
The equivalent control law (setting $\dot{s}_r=0$) is given by
\begin{equation}
\label{cntrlr}
u_{r_{eq}}=-r\dot{Z}_{re}-\textbf{P}_r\dot{Z}-\mathbf{R}_r+\ddot{Z}_{rd}
\end{equation}
The reachability control for the system \eqref{fast1} is
\begin{equation}
\label{reachr}
\bigtriangleup u_r=-\frac{\delta_r}{\epsilon_1} sgn(s_r)
\end{equation}
where, $\delta_r$ is a positive scalar. Then the sliding mode control law for inter group shape dynamics is written as
\begin{equation}
\label{totcntrlr}
F_{r}=u_{r_{eq}}+\bigtriangleup u_r
\end{equation}
\textit{Theorem 2} The control law \eqref{totcntrlr} will asymptotically stabilize the subsystem \eqref{eq102} in finite time.\\
\textit{Proof:} Define a Lyapunov function for the subsystem \eqref{eq102} as
\begin{equation}
\label{interliap}
V(s_r)=\frac{1}{2} s_r^Ts_r
\end{equation}
The time derivative of \eqref{interliap} gives
\begin{equation}
\dot{V}(s_r) = -\frac{\delta_r}{\epsilon_1}s_csgn(s_c) \leq-\frac{\delta_r}{\epsilon_1}\mid s_r \mid \leq-\frac{\delta_r}{\epsilon_1}V(s_r)^\frac{1}{2}
\end{equation}
As $\dot{V}(s_r)+\frac{\delta_r}{\epsilon_1}V(s_r)^\frac{1}{2}\leq 0$ it follows from Theorem 2 that the sliding surface $s_r$ is finite time stable and there exist a finite time $t_r\leq \frac{2\epsilon_1}{\delta_r}V(s_r(0))^{\frac{1}{2}}$ such that $z_{re}\rightarrow 0$ for all $t\geq t_r$.\\
\textit{3) Control law for intra group shape dynamics:} To serve the purpose of different time scale convergence, the control law of \eqref{eq101} is such chosen that error dynamics of the system \eqref{eq101} becomes singularly perturbed system as
\begin{equation}
\label{fast2}
\epsilon_1\epsilon_2\ddot{Z}_{se}=-\delta_ssgn(s_s)
\end{equation}
The sliding surface for the dynamics of \eqref{fast2} is chosen to be
\begin{equation}
\label{surfs}
s_s(t)=sZ_{se}+\dot{Z}_{se}
\end{equation}
The equivalent control law (setting $\dot{s}_s=0$)  is given by
\begin{equation}
\label{cntrls}
u_{s_{eq}}=-s\dot{Z}_{se}-\textbf{P}_s\dot{Z}-\mathbf{R}_s+\ddot{Z}_{sd}
\end{equation}
The reachability control for the system \eqref{fast2} is
\begin{equation}
\label{reachs}
\bigtriangleup u_s=-\frac{\delta_s}{\epsilon_1\epsilon_2}sgn(s_s)
\end{equation}
where, $\delta_s$ is a positive scalar. Then the sliding mode control law for inter group shape dynamics is written as
\begin{equation}
\label{totcntrls}
u_s=u_{s_{eq}}+\bigtriangleup u_s
\end{equation}
\textit{Theorem 3} The control law \eqref{totcntrls} will asymptotically stabilize the subsystem \eqref{eq101} in finite time.\\
\textit{Proof:} Define a Lyapunov function for the subsystem \eqref{eq101} as
\begin{equation}
\label{intraliap}
V(s_s)=\frac{1}{2} s_s^Ts_s
\end{equation}
The time derivative of \eqref{intraliap} gives
\begin{equation}
\dot{V}(s_s) = -\frac{\delta_s}{\epsilon_1\epsilon_2}s_ssgn(s_s) \leq-\frac{\delta_s}{\epsilon_1\epsilon_2}\mid s_s \mid \leq-\frac{\delta_s}{\epsilon_1\epsilon_2}V(s_s)^\frac{1}{2}
\end{equation}
As $\dot{V}(s_s)+\frac{\delta_s}{\epsilon_1\epsilon_2}V(s_s)^\frac{1}{2}\leq 0$ it follows from Theorem 3 that the sliding surface $s_s$ is finite time stable and there exist a finite time $t_s\leq \frac{2\epsilon_1\epsilon_2}{\delta_s}V(s_s(0))^{\frac{1}{2}}$ such that $z_{se}\rightarrow 0$ for all $t\geq t_s$.\\
\textbf{Remark:} As the settling times $t_c$, $t_r$, and $t_s$ depend on the initial values $V(s_c(0))$, $V(s_r(0))$, and $V(s_s(0))$ respectively and on the parameters $\delta_c$, $\delta_r$, and $\delta_s$ respectively, they can be selected such that $t_s>t_r>t_c$.

\section{Collision Avoidance}
The controllers of \eqref{totcntrlc}, \eqref{totcntrlr}, and \eqref{totcntrls} do not guarantee collision avoidance among the robots. Therefore, the barrier-like function of \cite{c13} is chosen as a potential function for collision avoidance. The modified form of the function for the robots $i,j\in \mathbb{N}$, $i,j=1,2,...,N$ is given by
\begin{equation}
\label{potf}
V_{ij}(p_i,p_j)=(p_i-p_j)\big ( b \  \text{exp} \big ( -\frac{\parallel p_i-p_j\parallel^2}{c}\big ) \big )
\end{equation}
where $b$, $c$ are positive constants and $p_i,p_j$ represents the position of $i$-th and $j$th robot respectively. Then the control input for the collision avoidance of $i$-th robot is the summation of all potential defined by \eqref{potf} of the robots $j$ inside the permissible distance $r$:
\begin{equation}
\label{pforce}
\bigtriangledown f_i=-\sum_{j=1,j\neq i}^{n} \frac{\partial V_{ij}(p_i,p_j)}{\partial p_i} ^T
\end{equation}
where $f_i\in\mathbb{R}^{2\times 1}$ and $\frac{\partial y}{\partial x}$ is the gradient of a scalar function $y$ (of dependent ($x$) and independent variables) with respect to $x$. Define a matrix $F\in\mathbb{R}^{2N\times 1}$ of control input based on avoidance potential of all robots $i=1,2,...,N$ as
\begin{equation}
\label{totf}
\bigtriangledown F=[\bigtriangledown f_1^T,\bigtriangledown f_2^T,...,\bigtriangledown f_N^T]^T
\end{equation}
To comply with the solutions of $\ddot{X}=-\bigtriangledown F$ under the transformation $Z=\Phi_M X$, define a vector of control input in the transformed domain as
\begin{equation}
\label{tp}
F_{pot}=\Phi_M \bigtriangledown F
\end{equation}
%\subsection{Case $1$: Three time scale analysis}
The vector $F_{pot}$ of \eqref{tp} is partitioned as $F_{pot}=[F_{pot\bold{s}}^T,F_{pot\bold{r}}^T,F_{pot\bold{c}}^T]^T$, where, $F_{pot\bold{s}}\in \mathbb{R}^{2\rho \times 1}$, $F_{pot\bold{r}}\in \mathbb{R}^{2(m-1)\times 1}$, and $F_{pot\bold{c}}\in \mathbb{R}^{2\times 1}$. Then the general sliding surface \cite{c13} for the intra and inter group and centroid dynamics given by
\begin{equation}
\label{smcca}
s_i(t)=iZ_{ie}+\dot{Z}_{ie}+F_{\textbf{i}pot}
\end{equation}
where, $i=c,r,s$. The equivalent control is then 
\begin{equation}
\label{smccaeq}
u_{i_{eq}}=-i\dot{Z}_{ie}-\textbf{P}_i\dot{Z}-\mathbf{R}_i+\ddot{Z}_{id}+\frac{d F_{\textbf{i}pot}}{d t}
\end{equation}
where, $i=c,r,s$. If the potential term of \eqref{smccaeq} is bounded, i.e. $\|\frac{d F_{\textbf{i}pot}}{dt} \|\leq \overline{F}_{\textbf{i}pot}$, for some known $\overline{F}_{\textbf{i}pot}$, $i=c,r,s$, then $\delta_i>\overline{F}_{\textbf{i}pot}+\gamma_i$ for the reachability control
\begin{equation}
\label{smccarc}
\bigtriangleup u_i=-\delta_isgn(s_i)
\end{equation}
where, $i=c,r,s$ and $\gamma_s\in\mathbb{R}^{2(n-m)}_{+}$, $\gamma_r\in\mathbb{R}^{2(n-m)}_{+}$, and $\gamma_c\in\mathbb{R}^{2}_{+}$ with $\mathbb{R}_{+}$ denotes positive real numbers. Then for a Lyapunov function $V_i=s_i^Ts_i$, the inequality reached from the control laws $F_i=u_{i_{eq}}+\bigtriangleup u_i$, $i=s,r,c$, is $\dot{V}_i=-\gamma_iV_i^{\frac{1}{2}}$ and therefore guarantees that the sliding manifold is reached in finite time $t_{max}^i=\frac{2V_i(0)}{\gamma_i}$ for $i=s,r,c$.

\section{Simulation Results}

\begin{figure}[h]
\centering
\includegraphics[width=8cm, height=6cm]{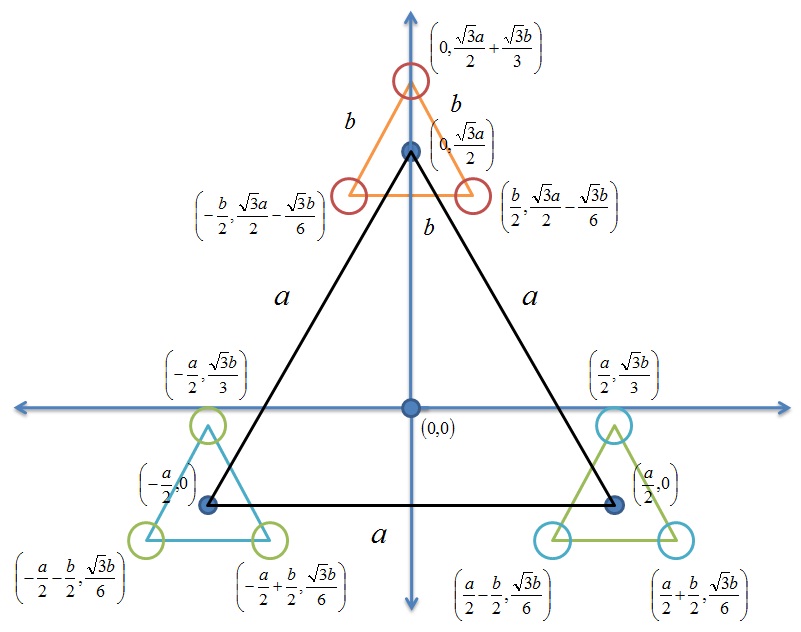}
\caption{Schematic Representation of Multiple Groups of Robots}
\label{fig:basis}
\end{figure}

\begin{figure}[h]
\centering
\includegraphics[width=8cm, height=5cm]{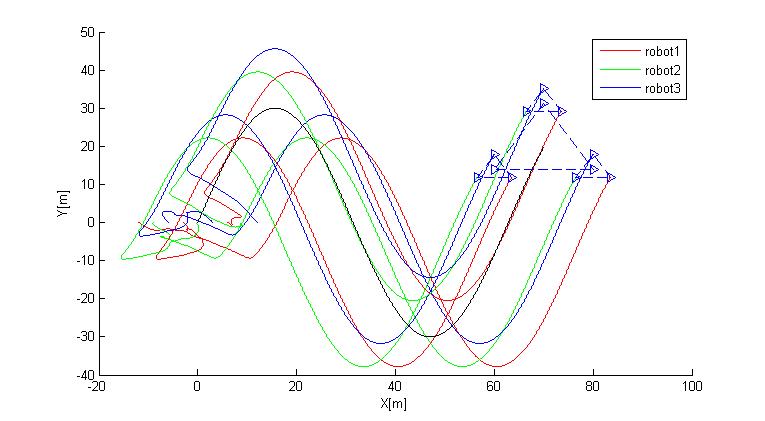}
\caption{Potential force based Formation control using transformation $\Phi_{M}$}
\label{fig:coll}
\end{figure}

\begin{figure}[h]
\centering
\includegraphics[width=8cm, height=6cm]{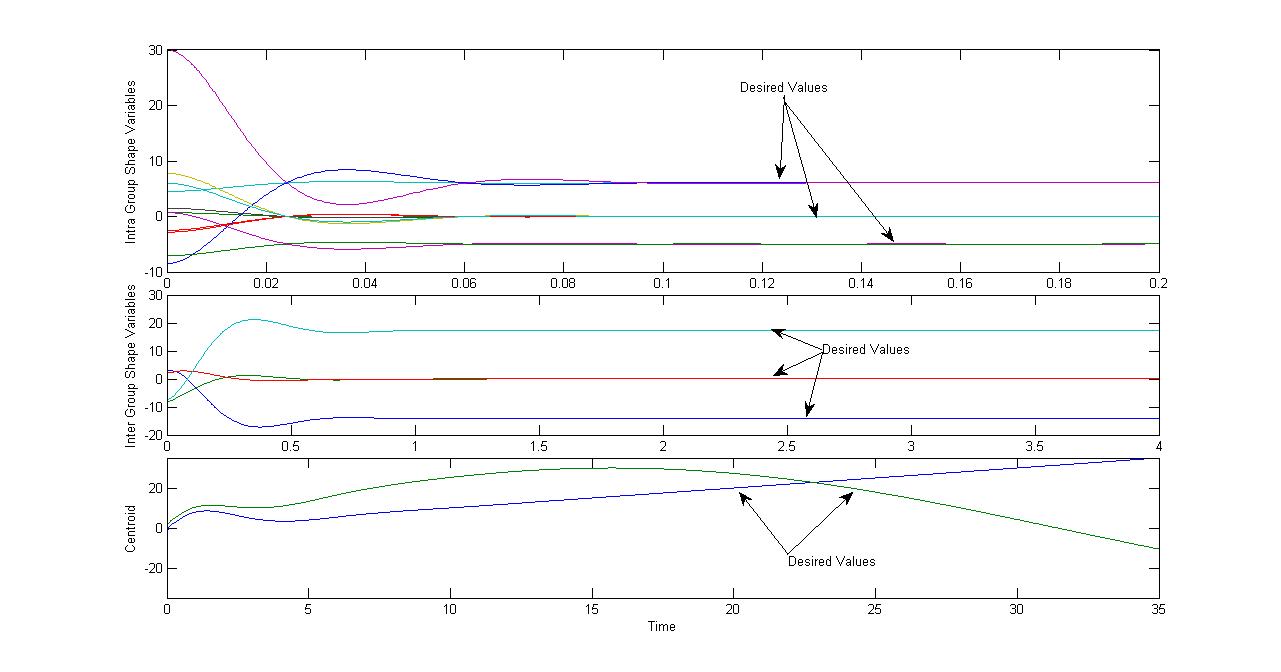}
\caption{Plot of intra and inter group shape variables and centroid vs time}
\label{fig:conv}
\end{figure}

We consider the dynamics of nonholonomic wheeled mobile robots of \eqref{801}. The system consists of $9$ such robots with $3$ robots in each of the $3$ groups as shown in Fig. \ref{fig:basis} with blue circles. $3$ robots in each group makes an equilateral triangle of side $b=7m$ and the centroids of each group forms an equilateral triangle of side $a=20m$ when connected.\\
Consider a planner formation with a formation basis defined as $\xi=[(\frac{a}{2}+\frac{b}{2},-\frac{\sqrt{3}a}{6}-\frac{\sqrt{3}b}{6}), (\frac{a}{2}-\frac{b}{2},-\frac{\sqrt{3}a}{6}-\frac{\sqrt{3}b}{6}), (\frac{a}{2},-\frac{\sqrt{3}a}{6}+\frac{\sqrt{3}b}{3}), (-\frac{a}{2}+\frac{b}{2},-\frac{\sqrt{3}a}{6}-\frac{\sqrt{3}b}{6}), (-\frac{a}{2}-\frac{b}{2},-\frac{\sqrt{3}a}{6}-\frac{\sqrt{3}b}{6}), (-\frac{a}{2},-\frac{\sqrt{3}a}{6}+\frac{\sqrt{3}b}{3}), (\frac{b}{2},\frac{\sqrt{3}a}{3}-\frac{\sqrt{3}b}{6}), (-\frac{b}{2},\frac{\sqrt{3}a}{3}-\frac{\sqrt{3}b}{6}), (0,\frac{\sqrt{3}a}{3}+\frac{\sqrt{3}b}{3})]^T$. The Jacobi vectors $Z$ for the Jacobi transformation $\Phi_M:X\rightarrow Z$ with $X=[p_{11}, p_{12}, p_{13}, p_{21}, p_{22}, p_{23}, p_{31}, p_{32}, p_{33}]^T$ is given below
\begin{equation}
\begin{cases}
Z_{1}\Rightarrow\begin{cases} 
z_{11}=\frac{1}{\sqrt{2}}(p_{12}-p_{11}) \\
z_{12}=p_{13}-\frac{1}{2}(p_{11}+p_{12}) \\
\end{cases}\\
Z_2\Rightarrow\begin{cases} 
z_{21}= \frac{1}{\sqrt{2}}(p_{22}-p_{21}) \\
z_{22}=p_{23}-\frac{1}{2}(p_{21}+p_{22}) \\
\end{cases}\\
Z_3\Rightarrow\begin{cases} 
z_{31}= \frac{1}{\sqrt{2}}(p_{32}-p_{31}) \\
z_{32}=p_{33}-\frac{1}{2}(p_{31}+p_{32}) \\
\end{cases}\\
Z_r\Rightarrow\begin{cases}
z_{r1}= \frac{1}{\sqrt{2}}(\mu_1-\mu_2) \\
z_{r2}=\mu_3-\frac{1}{2}(\mu_1+\mu_2) \\
\end{cases}\\
\end{cases}
\end{equation}
where, $\mu_1=\frac{1}{3}(p_{11}+p_{12}+p_{13})$, $\mu_2=\frac{1}{3}(p_{21}+p_{22}+p_{23})$, $\mu_3=\frac{1}{3}(p_{31}+p_{32}+p_{33})$.
The controller gain parameters are chosen as $s=r=c=1$, $\delta_s=\delta_r=\delta_c=1$ and $\epsilon_1=0.1$ and $\epsilon_2=0.1$. 
All the figures in this section show the trajectories of the robots moving in formation. The positions of the robots are marked by '$\triangleright$' and each group contains three robots marked with red, green and blue color. Potential force parameters are taken from \cite{c13}. 
The desired trajectory of the centroid of the formation is kept as $z_c=[t;30sin(0.1t)]$. 
\begin{figure}[h]
\centering
\includegraphics[width=8cm, height=5cm]{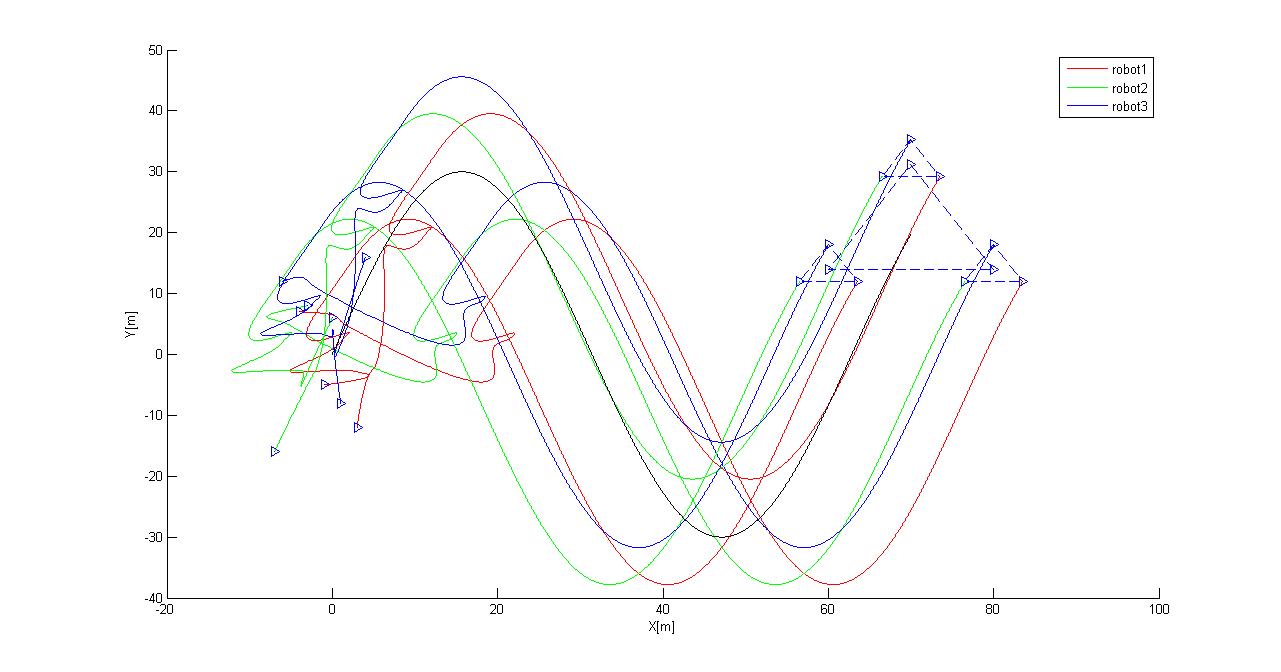}
\caption{Formation control using transformation $\Phi_{M}$}
\label{fig:wcoll}
\end{figure}
In Fig. \ref{fig:wcoll}, it is shown that the robots converge to the desired formation. Potential force has not been considered for the simulation in Fig. \ref{fig:wcoll}. The convergence of robots to the desired formation with collision avoidance, is depicted in Fig. \ref{fig:coll}.\\
Fig. \ref{fig:conv} shows the convergence time of the states in the transformed domain separately (without applying potential force). All the intra group shape variables $Z_1 \ldots Z_6$ converge faster than inter group shape variables $Z_7$ and $Z_8$. It can also be seen from the figures, that the convergence of the centroid is the slowest of all. It is evident from Fig. \ref{fig:conv} that the intra group shape variables converge to desired value at $t=0.1sec$. The inter group shape variables converge at time $t=1sec$ and the trajectory of centroid converges to the desired value at $t=10sec$. Thus convergence of intra group shape variables are $10$ times faster than the convergence of inter group shape variables. Again, convergence of the trajectory of centroid is $10$ times faster than the convergence of inter group shape variables.

\section{Conclusion}
This paper addresses the design of sliding mode controller for multiple groups of robots under linear transformation. We first give a linear transformation for multiple groups of robots using Jacobi transformation for single group of robots, to integrate the nonholonomic dynamics of $n$ robots undergoing planner formation. We name it \textit{Centroid Based Transformation}. The transformation separates the combined dynamics of $n$ robots into intra and inter group shape dynamics and the dynamics of the centroid. The parameters of the sliding mode controller is such chosen that the closed loop dynamics becomes singularly perturbed system. In effect different dynamics reaches different sliding surfaces at different \textit{finite} time. For collision avoidance, negative gradient of repulsive potential function of \cite{c13} has also been appended the proposed feedback controller. A sliding surface is chosen such that the collision avoidance controller reach sliding surface in finite time. Simulation results show the effectiveness of our proposed controllers.

\end{document}